\documentclass[prb,aps,twocolumn,amsmath,amssymb,floatfix,showpacs,
superscriptaddress]{revtex4}

\usepackage[dvips]{graphics}
\usepackage{color}
\definecolor{dred}{rgb}{0,0,0.6}

\begin{document}

\title{Externally controlled selective spin transfer through a 
two-terminal bridge setup}

\author{Santanu K. Maiti}

\email{santanu.maiti@isical.ac.in}

\affiliation{Physics and Applied Mathematics Unit, Indian Statistical
Institute, 203 Barrackpore Trunk Road, Kolkata-700 108, India}

\begin{abstract}

A new way of getting controlled spin dependent transport through a 
two-terminal bridge setup is explored. The system comprises a 
magnetic quantum ring which is directly coupled to a magnetic quantum 
wire and subjected to an in-plane electric field perpendicular to the 
wire. Without directly changing system parameters, one can regulate spin 
currents simply by tuning the external electric field under a finite bias
drop across the wire. For some particular field strengths a high degree 
of spin polarization can be achieved and thus the system can essentially 
be utilized as an externally controlled spin polarized device. A detailed
comparison of spin current magnitudes obtained from other bridge setups is
also examined to make the present investigation a self contained study.

\end{abstract}


\maketitle

\section{Introduction}

The phenomenon of spin polarized transport~\cite{sp1,sp2,sp3} in 
low-dimensional systems has emerged as one of the most promising area over 
the last few decades in condensed matter physics due to its potential 
application in nano-science and technology~\cite{nano1,nano2}. The 
advancement of nano-lithographic techniques along with sophisticated 
instrumentation facilities have enabled experimentalists to explore spin 
dependent transport through different tailor made geometries and test 
their ability in the realization of future nano-scale spin based electronic 
devices~\cite{ex1,ex2,ex3}. With the discovery of giant magneto-resistance 
(GMR) effect~\cite{gmr} in Fe/Cr magnetic multilayers during 1980's a new 
branch in condensed matter physics, the so-called {\em spintronics}, has 
been developed which deals with the possibilities of exploring electron 
spin in transport properties. This phenomenon has lead to the revolutionary 
progress in devices making, data processing, quantum computations and many 
others~\cite{dev1,dev2,dev3,dev4}. Three most fundamental steps are 
involved~\cite{spinsteps1,spinsteps2} in designing spin based electronic 
devices those are: injection of spin through interfaces, propagation of 
spin through material and finally the detection of spin. Quantum confined 
nanostructures are the ideal candidates for it since they have considerably
large spin coherence time. Therefore, the studies associated with spin 
dependent transport in nanostructures are of great importance from the 
aspect of theoretical understanding as well as technological progress,
especially to design spintronic devices. 

In order to design a controllable spintronic device, the most crucial
requirement is the generation of polarized spin currents and proper 
regulation of these currents. Therefore, modeling of spin filter is of 
great importance. Over the last few years many theoretical~\cite{th1,th2,
sm1,th3,sm2,th4,thnew1,thnew2,thnew3,thnew4,th5,th6,sm3,th7,th8} as well 
as experimental~\cite{ex1,ex2,ex3,expt} 
works have been done to explore spin dependent transport at 
nano-scale level and to design efficient spin filter with higher spin 
polarizability. One common route of developing a spin filter is by using 
ferromagnetic electrodes~\cite{trend1,trend2} though its experimental 
realization is somewhat complicated since spin injection from these 
electrodes becomes quite difficult due to large resistivity mismatch. 
Keeping in mind the above issue, a large section of the existing literature 
rather suggests to design spin filter device using the intrinsic properties 
of materials, for example, spin-orbit (SO) interactions~\cite{intrinsic1,
intrinsic2,intrinsic3,intrinsic4,intrinsic5,intrinsic6}. Usually two types 
of SO interactions, Rashba and Dresselhaus, are encountered in solid state 
materials~\cite{rashba,dressel,winkler,skm} depending on their sources. 
The Rashba SO coupling in a material is attributed to an electric field 
that originates from the lacking of structural symmetry whereas, the 
Dresselhaus SO interaction appears from
\begin{figure}[ht]
{\centering \resizebox*{6.8cm}{4.5cm}{\includegraphics{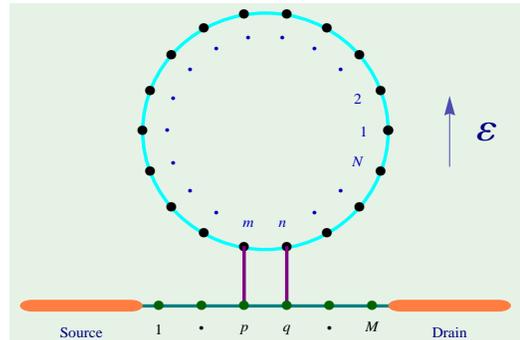}}\par}
\caption{(Color online). A magnetic quantum wire, sandwiched between two
one-dimensional non-magnetic source and drain electrodes, is directly 
coupled to a magnetic quantum ring. An in-plane electric field 
$\mathcal{E}$, perpendicular to the wire, is applied to the ring.}
\label{model}
\end{figure}
the bulk inversion asymmetry. Out of these two, Rashba SO interaction has 
been attracted much attention in the field of spintronics since its coupling 
strength can be tuned by electrostatic means i.e., applying external gate
voltages~\cite{gt1,gt2,gt3,gt4,gt5} which provides controlled spin dependent 
transport and manipulates electronic spin state. In presence of SO coupling 
polarized spin currents in output terminals of a {\em multi-terminal} 
conductor can be achieved from a completely unpolarized electron beam 
injected to its input terminal. To date, many works have been 
done~\cite{mlt1,mlt2,mlt3,mlt4,mlt5,skm3} considering different 
multi-terminal geometries. For example, in 2006, Peeters {\em et al.} have 
analyzed how a simple mesoscopic ring with one input and two output terminals 
can be utilized as an electron spin beam splitter~\cite{mlt1} in presence 
of Rashba SO coupling. In other work, Kislev and Kim have proposed that a 
planar T-shaped geometry with a ring resonator provides~\cite{mlt2} Rashba 
SO interaction induced polarized spin currents in output terminals. Among 
these, many other groups~\cite{mlt3,mlt4} have also put forward different 
key ideas in this particular realm. But, it should be stressed, unlike 
multi-terminal bridge systems, only SO coupling is incapable of producing 
polarized spin current when a sample is coupled to single input and a single 
output lead i.e., in a two-terminal bridge setup~\cite{mlt2}.
In presence of SO coupling the time-reversal symmetry gets preserved, and 
hence, it doesn't break Kramer's degeneracy between $|k\uparrow\rangle$ 
and $|-k\downarrow\rangle$ states which results vanishing spin current in the 
output lead. This degeneracy is broken when the material is subjected to an 
external magnetic field, and under this situation a two-terminal conductor 
subjected to SO interaction can exhibit~\cite{raba,smpola} polarized spin 
current in its output lead. But, this approach is not quite suitable since 
confining a strong magnetic field in a small region like a quantum dot (QD) 
or a quantum ring (QR) is extremely difficult. Therefore, further studies 
are still required to develop a possible route of getting controlled spin 
selective transmission in a two-terminal geometry.

In the present paper, we propose a theoretical model to realize spin 
selective transmission through a two-terminal conducting bridge and explore 
the possibilities to control spin dependent currents {\em externally} 
without directly changing the physical parameters of the system. The model 
quantum system is designed by a magnetic quantum wire (MQW), sandwiched 
between two non-magnetic (NM) source and drain electrodes, which is again
directly coupled to a magnetic quantum ring (MQR). The ring is subjected to 
an external electric field and it is the key controlling parameter of our 
present investigation. The main motivation behind the 
consideration of this particular geometry is to investigate the interplay
of the MQR, which behaves like a correlated disordered ring in presence of
external electric field (will be discussed later in the appropriate 
sub-section), and the MQW to achieve selective spin transport in a 
two-terminal junction. It is well known that in any magnetic nanostructure
there is always a band misalignment between up and down spin electrons
irrespective of any external electric field, and therefore, one can get
pure spin current upon selecting the Fermi energy to a suitable energy 
zone. But, for such a nanostructure neither spin current can be controlled
efficiently nor large spin current can be achieved. To circumvent these 
issues we propose a new model, shown in Fig.~\ref{model}, where spin 
current can be tuned systematically by means of external electric field
and controlling this field large current can be achieved. We strongly 
believe that the design of such a system is of great concern in the 
current era of nanofabrication. Within a 
tight-binding (TB) framework and based on Green's function formalism
we show that selective spin currents are available at the output terminal 
and their magnitudes can be regulated by means of external in-plane electric
field. This system also exhibits a high degree of spin polarization for some
typical field strengths. Our theoretical results promote practical 
applications of externally controlled spin polarized quantum devices.
Finally, to substantiate the proposed system as an 
efficient two-terminal externally controlled spin-filter device, here 
we also compare the spin current magnitudes considering other geometrical 
systems. Analyzing the results we ensure that the model presented in
Fig.~\ref{model} is the most suitable one.

The rest of the paper is organized as follows. In Section II we describe 
the model together with theoretical formulations for the calculations. 
Essential findings are described in Section III. Finally, we summarize 
our results in Section IV.

\section{Model and theoretical framework}

\subsection{Model and Hamiltonian}

Let us begin by referring to Fig.~\ref{model} where a MQW, coupled to a 
MQR, is sandwiched between two semi-infinite one-dimensional NM electrodes 
commonly known as source and drain. The ring is subjected to an in-plane 
electric field $\mathcal{E}$, perpendicular to the wire, which controls 
selective spin transmission across this two-terminal junction. To emphasize 
the effect of quantum interference on electronic transport we connect the 
ring to the wire through two vertical bonds, instead of attaching them via 
a single bond. 

Using a tight-binding approach we describe the model quantum system and in
the absence of any electron-electron (e-e) interaction this scheme is 
extremely suitable for analyzing electron transport through a conducting 
bridge~\cite{tb1,tb2,tb3,tb4,tb5,tb6,tb7,tb8}. The single particle TB 
Hamiltonian that includes the MQW, MQR and NM source and drain electrodes 
can be written as,
\begin{equation}
H=H_{\mbox{\tiny c}} + H_{\mbox{\tiny el}} + H_{\mbox{\tiny tn}}
\label{equ1}
\end{equation}
where three different terms in the right side represent three distinct 
regions of the complete system. These terms are elaborately explained 
as follows. 

The first term $H_{\mbox{\tiny c}}$ corresponds to the Hamiltonian of 
the conductor within the electrodes i.e., the ring including the wire. 
Each site of the ring as well as the wire is associated with a local
magnetic moment with amplitude $h_i$ (say, for $i$-th site) and the 
orientation of such a magnetic moment is specified by the polar angle 
$\theta_i$ and azimuthal angle $\phi_i$ in spherical polar co-ordinate 
system. The 
orientations of these local moments can be controlled by applying a 
magnetic field. Under nearest-neighbor hopping approximation the TB 
Hamiltonian $H_{\mbox{\tiny c}}$ becomes,
\begin{eqnarray}
H_{\mbox{\tiny c}} & = & \sum_i \mbox{\boldmath $c$}_i^{r\dagger} 
\left(\mbox{\boldmath $\epsilon$}_i^r -
\vec{\mbox{\boldmath $h$}}_i.\vec{\sigma} \right) \mbox{\boldmath $c$}_i^r + 
\sum_i \left(\mbox{\boldmath $c$}_{i+1}^{r\dagger} \mbox{\boldmath $t$}_r
\mbox{\boldmath $c$}_i^r + h.c. \right) 
\nonumber \\
 & + & \sum_i \mbox{\boldmath $c$}_i^{w\dagger} 
\left(\mbox{\boldmath $\epsilon$}_i^w -
\vec{\mbox{\boldmath $h$}}_i.\vec{\sigma} \right) \mbox{\boldmath $c$}_i^w +
\sum_i \left(\mbox{\boldmath $c$}_{i+1}^{w\dagger} \mbox{\boldmath $t$}_w
\mbox{\boldmath $c$}_i^w + h.c. \right)
\nonumber \\
 & + & \left(\mbox{\boldmath $c$}_m^{r\dagger} 
\mbox{\small \boldmath $\lambda$} \mbox{\boldmath $c$}_p^w + 
\mbox{\boldmath $c$}_n^{r\dagger} \mbox{\small \boldmath $\lambda$}
\mbox{\boldmath $c$}_q^w + h.c. \right)
\label{equ2}
\end{eqnarray}
where, \\
$\mbox{\boldmath $c$}_i^{r\dagger} =\left(\begin{array}{cc}
c_{i \uparrow}^{r\dagger} & c_{i \downarrow}^{r\dagger} \end{array}\right);$
$\mbox{\boldmath $c$}_i^r =\left(\begin{array}{c}
c_{i \uparrow}^r \\
c_{i \downarrow}^r \end{array}\right);$
$\mbox{\boldmath $c$}_i^{w\dagger} =\left(\begin{array}{cc}
c_{i \uparrow}^{w\dagger} & c_{i \downarrow}^{w\dagger} \end{array}\right);$
$\mbox{\boldmath $c$}_i^w =\left(\begin{array}{c}
c_{i \uparrow}^w \\
c_{i \downarrow}^w \end{array}\right);$
$\mbox{\boldmath $\epsilon$}_i^r=\left(\begin{array}{cc}
\epsilon_i^r & 0 \\
0 & \epsilon_i^r \end{array}\right);$ 
$\mbox{\boldmath $\epsilon$}_i^w=\left(\begin{array}{cc}
\epsilon_i^w & 0 \\
0 & \epsilon_i^w \end{array}\right);$ \\
$\mbox{\boldmath $t$}_r=t_r\left(\begin{array}{cc}
1 & 0 \\
0 & 1 \end{array}\right);$
$\mbox{\boldmath $t$}_w=t_w\left(\begin{array}{cc}
1 & 0 \\
0 & 1 \end{array}\right);$
$\mbox{\small \boldmath $\lambda$}=$\mbox{\small $\lambda$}$
\left(\begin{array}{cc}
1 & 0 \\
0 & 1 \end{array}\right);$
$\vec{\mbox{\boldmath $h$}}_i.\vec{\sigma}= h_i\left(\begin{array}{cc}
\cos \theta_i & \sin \theta_i e^{-j \phi_i} \\
\sin \theta_i e^{j \phi_i} & -\cos \theta_i \end{array}\right).$ \\
~\\
In the above expression (Eq.~\ref{equ2}), the 1st and 2nd terms are
associated with the magnetic quantum ring of $N$ atomic sites, whereas 
for the magnetic wire containing $M$ atomic sites the 3rd and 4th terms 
are used, and the last term describes the coupling between them. 
$c_{i\sigma}^{r\dagger}$ and $c_{i\sigma}^r$ are the
creation and annihilation operators, respectively, for an electron
with spin $\sigma(\uparrow,\downarrow)$ at the site $i$ of the ring,
while for the wire they are represented by $c_{i\sigma}^{w\dagger}$ and 
$c_{i\sigma}^w$, respectively. $\epsilon_i^r$ gives the site energy
and $t_r$ corresponds to the nearest-neighbor hopping integral in the 
ring. Similarly, for the wire they are respectively described by 
$\epsilon_i^w$ and $t_w$. The factor $\vec{\mbox{\boldmath 
$h$}}_i.\vec{\sigma}$ describes interaction of the spin of injected 
electron to the local magnetic moment placed at $i$-th site.
In order to elucidate the role of quantum interference
on electronic conduction, MQR is attached to the MQW by more than a 
single interaction, as shown in Fig.~\ref{model}. Any two atomic sites
$m$ and $n$ (not necessarily nearest-neighbor) of the MQR can be connected
to the atomic sites $p$ and $q$ of the MQW by two vertical lines to get 
two different connecting paths between the MQR and MQW. As the essential 
features of our present investigation can be acquired considering
$m$ and $n$ as nearest-neighbor sites (the simplest configuration), we 
couple the site $m$ of the MQR to the site $p$ of the MQW by a single 
bond, and similarly, the site $n$ is connected to the site $q$ by another
bond (Fig.~\ref{model}, for this configuration $p$ and $q$ are also the
nearest-neighbor sites). The hopping integral between the sites $m$ and 
$p$ is described by the parameter $\mbox{\small $\lambda$}$, and for the
other two sites $n$ and $q$ it is also characterized by 
$\mbox{\small $\lambda$}$. The main target of this particular geometry
is to find the selective and controlled spin transmission and the 
interplay of energy levels of the ring which is coupled to a quantum 
wire in presence of a finite bias. This can essentially be done with the
help of external electric field which regulates on-site potentials of MQR
upon the variation of electric field. In presence of this field, site 
energy of the MQR becomes field dependent and doing some simple and 
straight-forward mathematical steps one can get the site energy for a 
$N$-site ring as: $\epsilon_i^r=(e a N\mathcal{E}/2\pi)\cos[2\pi(i-1)/N]$, 
where $e$ gives the electronic charge, $a$ corresponds to the lattice 
spacing and $\mathcal{E}$ measures the electric field strength. This 
relation can be simplified by introducing the dimensionless electric 
field strength $\xi$ as $\epsilon_i^r=\left(Nt_w\xi/2\pi \right) 
\cos[2\pi(i-1)/N]$, where $\xi=e a \mathcal{E}/t_w$. In 
absence of any electric field, local on-site energy $\epsilon_i^r$ of the 
ring becomes constant, and therefore, we can fix it at zero without loss 
of any generality. This is exactly what we get from the above relation.

The side attached electrodes are assumed to be semi-infinite, non-magnetic 
and free from any kind of impurities. We can express them like,
\begin{equation}
H_{\mbox{\tiny el}} = \sum_{\alpha} H_{\alpha} 
\label{equ3}
\end{equation}
where $\alpha=S$ and $D$ for the source and drain, respectively. In TB 
framework $H_{\alpha}$ reads as,
\begin{equation}
H_{\alpha} = \sum \limits_i \mbox{\boldmath$d$}_i^{\dagger} 
\mbox{\boldmath$\epsilon$}_l^{\alpha} \mbox{\boldmath$d$}_i + 
\sum_i \left(\mbox{\boldmath$d$}_i^{\dagger} \mbox{\boldmath$t$}_l^{\alpha}
\mbox{\boldmath$d$}_{i+1} + h.c. \right)
\label{equ4}
\end{equation}
with
$\mbox{\boldmath $\epsilon$}_l^{\alpha}=\epsilon_l^{\alpha}
\left(\begin{array}{cc}
1 & 0 \\
0 & 1 \end{array}\right)$ and
$\mbox{\boldmath $t$}_l^{\alpha}=t_l^{\alpha}
\left(\begin{array}{cc}
1 & 0 \\
0 & 1 \end{array}\right)$.\\
~\\
$\epsilon_l^{\alpha}$ and $t_l^{\alpha}$ are the site energy and
nearest-neighbor hopping integral, respectively, in the $\alpha$-th lead,
and $d_{i\sigma}^{\dagger}$ ($d_{i\sigma}$) is the creation (annihilation)
operator of an electron with spin $\sigma$ at $i$-th site of the electrodes.
These electrodes are coupled through the atomic sites $1$ and $M$ of the
wire via the coupling parameter $t_c$. Following the same prescription the 
wire-to-lead coupling Hamiltonian gets the form,
\begin{equation}
H_{\mbox{\tiny tn}} = \mbox{\boldmath$c$}^{w\dagger}_1
\mbox{\boldmath$t$}_c \mbox{\boldmath$d$}_0 +
\mbox{\boldmath$d$}_0^{\dagger} \mbox{\boldmath$t$}_c \mbox{\boldmath$c$}^w_1
+ \mbox{\boldmath$c$}_M^{w\dagger} \mbox{\boldmath$t$}_c 
\mbox{\boldmath$d$}_{M+1} + \mbox{\boldmath$d$}_{M+1}^{\dagger}
 \mbox{\boldmath$t$}_c \mbox{\boldmath$c$}^w_M
\label{equ5}
\end{equation}
with
$\mbox{\boldmath $t$}_c=t_c\left(\begin{array}{cc}
1 & 0 \\
0 & 1 \end{array}\right).$

\subsection{Transmission probability, junction current and spin polarization 
coefficient: Green's function approach}

To calculate spin dependent transmission probabilities, junction currents 
and spin polarization coefficient we use Green's function 
formalism~\cite{datta1,datta2}. In this approach, transmission probability 
$T_{\sigma \sigma^{\prime}}$ of an injecting electron with spin $\sigma$ 
which gets transmitted through the drain electrode with spin 
$\sigma^{\prime}$ is written as~\cite{datta1,datta2}
$T_{\sigma \sigma^{\prime}}=\mbox{Tr}\left[\Gamma_{\mbox{\tiny S}}^{\sigma} 
G_c^r \Gamma_{\mbox{\tiny D}}^{\sigma^{\prime}} G_c^a\right]$. When
$\sigma=\sigma^{\prime}$ we get pure spin transmission, while for the other 
case ($\sigma \ne \sigma^{\prime}$) spin flip transmission is obtained. 
$G_c^r$ and $G_c^a$ are the retarded and advanced Green's functions, 
respectively, of the conductor i.e., MQR including the MQW sandwiched 
between the electrodes. 
$G_c^r=\left(E-H_{\mbox{\tiny c}} - \sum\limits_{\sigma}
\Sigma_{\mbox{\tiny S}}^{\sigma} - \sum\limits_{\sigma}
\Sigma_{\mbox{\tiny D}}^{\sigma}\right)^{-1}$, where $E$ is the energy of 
an injecting electron, and $\Sigma_{\mbox{\tiny S}}^{\sigma}$ and
$\Sigma_{\mbox{\tiny D}}^{\sigma}$ are the self-energies due to coupling 
of the MQW to the electrodes and $\Gamma_{\mbox{\tiny S}}^{\sigma}$ and 
$\Gamma_{\mbox{\tiny D}}^{\sigma}$ are their imaginary parts. For 
comprehensive derivations of these self-energy matrices, go through the 
references~\cite{datta1,datta2}. In these pioneering 
references it is shown that the self-energy can be expressed as a linear 
combination of real and imaginary parts, where the real part measures
the shift of energy levels, while the other part gives the broadening 
of these levels. The finite imaginary part appears due to incorporation 
of the semi-infinite electrodes having continuous energy spectrum.

The spin dependent current $I_{\sigma \sigma^{\prime}}$ passing through
the junction can be obtained from the Landauer-B\"{u}ttiker formalism. 
It is written as~\cite{datta1,datta2},
\begin{equation}
I_{\sigma \sigma^{\prime}} (V)= \frac{e}{h} 
\int \left[f_{\mbox{\tiny S}}(E)-f_{\mbox{\tiny D}}(E)\right] 
T_{\sigma \sigma^{\prime}}(E) \, dE
\label{equ6}
\end{equation}
where, $f_{\mbox{\tiny S}}(E)$ and $f_{\mbox{\tiny D}}(E)$ are the Fermi 
distribution functions of the source and drain with electro-chemical 
potentials $\mu_{\mbox{\tiny S}}$ ($=E_F+eV/2$) and $\mu_{\mbox{\tiny D}}$
($=E_F-eV/2$), respectively. $E_F$ gives the equilibrium Fermi energy and 
it can be controlled via external gate voltages. From 
Eq.~\ref{equ6} we can evaluate pure spin currents (up spin electron gets 
transmitted as up spin, and similarly for down spin electron which is
transferred as a down spin) as well as spin flip currents (up spin electron 
gets flipped when it reaches to the drain through the bridging magnetic 
conductor and vice versa) by integrating proper transmission coefficients 
over a particular voltage window, and eventually, we obtain the net up and 
down spin currents.
These are: $I_{\uparrow}=I_{\uparrow \uparrow} + I_{\downarrow \uparrow}$
and $I_{\downarrow}=I_{\downarrow \downarrow} + I_{\uparrow \downarrow}$.

Finally, spin polarization coefficient of total current is measured from 
the relation~\cite{rai,nmn},
\begin{eqnarray}
P &=& \left|\frac{I_{\uparrow} - I_{\downarrow}}{I_{\uparrow} 
+ I_{\downarrow}}\right| \nonumber \\
 &=& \left|\zeta_{\uparrow}-\zeta_{\downarrow}\right|
\label{equ7}
\end{eqnarray}
where, $\zeta_{\sigma}=I_{\sigma}/(I_{\uparrow}+I_{\downarrow})$ 
describes the spin filter efficiency. The quantities $I_{\uparrow}$ and 
$I_{\downarrow}$ can also be derived directly from Eq.~\ref{equ6} by 
integrating the net up and down spin transmission probabilities those 
are respectively expressed as
$T_{\uparrow}=T_{\uparrow \uparrow} + T_{\downarrow \uparrow}$ and
$T_{\downarrow}=T_{\downarrow \downarrow}+T_{\uparrow \downarrow}$. In our 
theoretical description all the mathematical expressions are framed 
considering the quantization direction along the positive z-axis where
$\sigma_z$ gets the form:
$$\sigma_z=\left(\begin{array}{cc}
1 & 0 \\
0 & -1 \end{array}\right).$$ 

\section{Numerical results and discussion}

\begin{figure*}
{\centering \resizebox*{14cm}{17.5cm}{\includegraphics{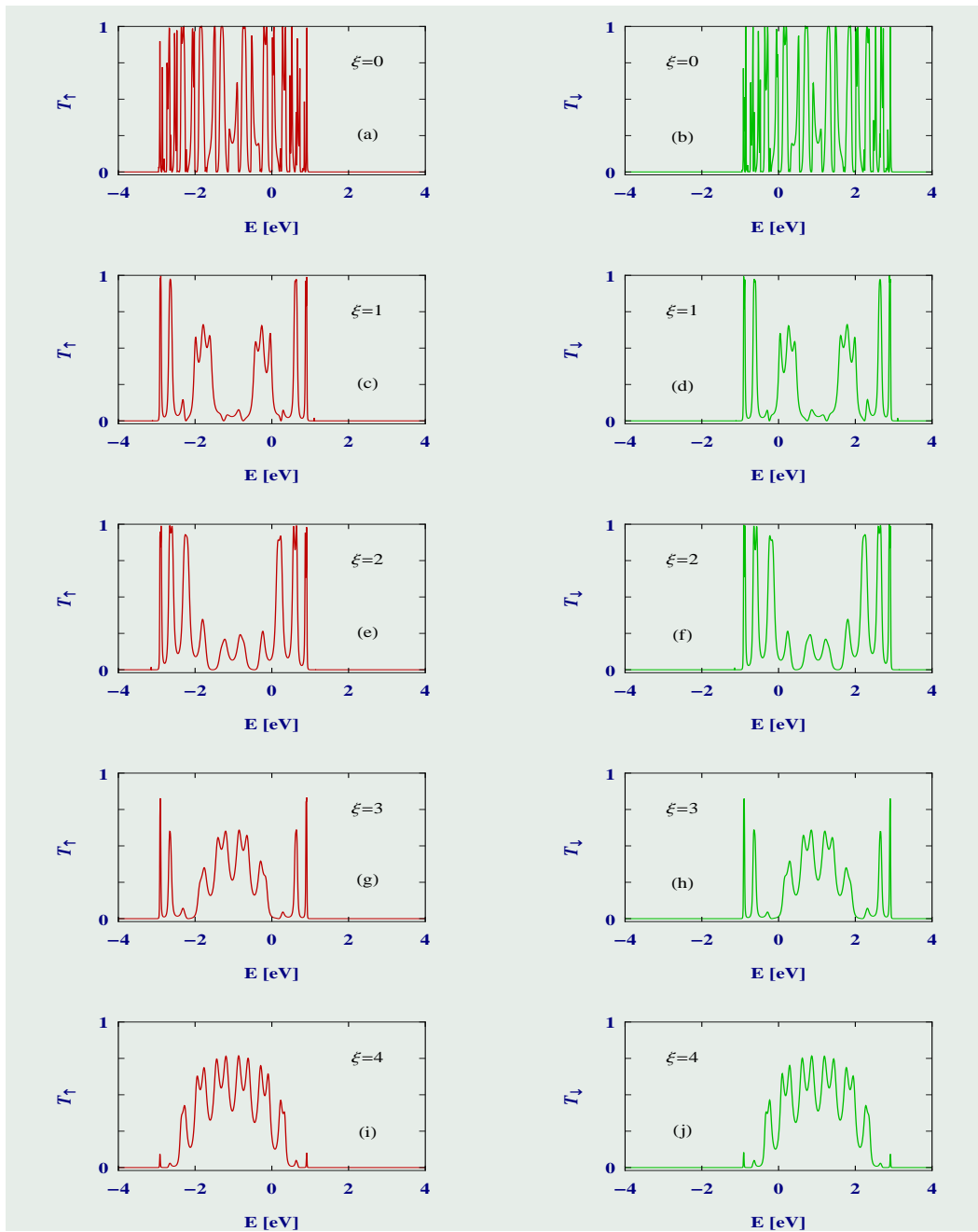}}\par}
\caption{(Color online). Energy dependence of $T_{\uparrow}$ and 
$T_{\downarrow}$ for different values of dimensionless electric field 
strength $\xi$. The other physical parameters are: $N=42$, $M=20$, $m=32$, 
$n=33$, $p=10$, $q=11$ and $\lambda=1$ eV.}
\label{updn}
\end{figure*}
According to the above theoretical formulation, described in Sec. II, we 
are now ready to present our numerical results for spin dependent transmission 
probabilities and spin polarization coefficient, and, the effect of in-plane
electric field on them. During calculations we fix the electronic temperature
of the system to zero. The other common parameters are chosen as follows. 
Both in the MQW and MQR we assume that all the magnetic moments are aligned 
along positive z-axis i.e., $\theta_i=0$ and $\phi_i=0$ and they are equal 
in magnitude ($h_i=1$ eV for all the magnetic sites $i$). 
The site energies in the electrodes ($\epsilon_l^{\alpha}$)  
and in the magnetic wire ($\epsilon_i^w$) are set to zero. For the ring, 
the site energies ($\epsilon_i^r$) are no longer identical since they are 
field dependent for non-zero electric field as prescribed in our theoretical 
description. The hopping integrals $t_r$, $t_w$ and $t_c$ are set to $1$ eV, 
whereas the hopping integral in the electrodes $t_l^{\alpha}$ is fixed at 
$2$ eV. Finally, we set the lattice spacing $a=1\, A^{\circ}$.

\subsection{Two-terminal transmission coefficients}

We start by analyzing the influence of in-plane electric field on  
transmission probabilities. The results for net up ($T_{\uparrow}$) and 
down ($T_{\downarrow}$) spin transmission probabilities as a function of 
injecting electron energy $E$ are depicted in Fig.~\ref{updn}, where  
sizes of the MQR and MQW are chosen as $N=42$ and $M=20$, respectively, 
and the other physical parameters are set at $m=32$, $n=33$, $p=10$,
$q=11$ and $\lambda=1$ eV. The transmission spectra exhibit several 
interesting patterns both for up and down spin electrons which are 
analyzed as follows. In
absence of external electric field the transmission coefficients 
$T_{\uparrow}$ and $T_{\downarrow}$ provide sharp resonant peaks 
(see Figs.~\ref{updn}(a) and (b)) associated with energy eigenvalues of 
the conductor, and for most of these resonant energies the transmission
probability reaches very close to unity. The transmittance spectrum gets 
significantly modified with external electric field and depending on its
strength, low and high, two anomalous features are obtained. At lower 
value of $\xi$, say $\xi=1$, resonant peaks are broadened and they are 
separated with non-uniform energy gaps (see  Figs.~\ref{updn}(c) and (d)). 
In addition, the heights of some of these resonant peaks are also suppressed 
compared to the electric field free case, which is noticed by comparing the 
spectra shown in the top two rows of Fig.~\ref{updn}. With increasing the 
field strength, say $\xi=2$, some resonant peaks with larger widths 
(Figs.~\ref{updn}(e) and (f)) are generated across the edges of allowed 
energy band, but around the energy band centre height of the peaks is 
reduced enormously. If the field strength is increased further, the 
features described above get reversed. More resonant peaks appear around 
the energy band centre with increasing heights (Figs.~\ref{updn}(g) and 
(h)) and for large enough field strength gapless spectrum is visible 
(Figs.~\ref{updn}(i) and (j)).

Now we try to explain these spectral features physically. The transmission 
spectrum of a bridge system is directly associated with eigenenergies of the 
conductor clamped between two electrodes. In absence of any electric field,
the conductor within the electrodes behaves like a perfect one since
site energies of both the MQR and MQW are identical. For such a perfect
conductor the energy levels are conducting in nature and all of them 
contribute to the electronic transmission which results a large number of
resonant peaks in $T_{\sigma}$-$E$ spectrum. For non-zero electric field,
site energies of the MQR are no longer identical to the MQW since they 
are now field dependent and none of them are equal in magnitude. Under
this situation the MQR is treated as a correlated disordered ring and 
hence the combined system (MQW including MQR) within the electrodes can
be called as an ordered-disordered coupled conductor. In a fully disordered 
system where all site energies are different localized energy states are
expected and they become more localized with increasing the disorderness. 
While, for an ordered-disordered coupled
system a set of conducting states together with localized energy levels
are obtained and these conducting states become less conducting with
increasing disorderness in the weak disorder regime since these two regions
are coupled with each other. The situation is somewhat different in the 
limit of strong disorder. In this limit, the ordered and disordered regions 
are almost decoupled from each other, and accordingly, the conducting states 
which arise from the perfect region i.e., MQW are influenced very weakly by 
the localized states generated from the MQR. With these peculiar features of
energy eigenstates in an ordered-disordered coupled system, depending on
the strength of disorderness associated with in-plane electric field, the 
characteristics properties of $T_{\sigma}$-$E$ shown in Fig.~\ref{updn} can 
be easily understood. For the lower field strength, less conducting states 
those are affected by the disordered region contribute to the electronic 
conduction providing few resonant peaks with reduced amplitudes in the 
$T_{\sigma}$-$E$ spectrum. On the other hand, for large enough electric 
field electrons get transmitted only through the perfect region (MQW), 
and therefore, a gapless spectrum with larger amplitude is obtained.

In addition to the above facts it is interesting to note that the up and 
down spin electrons are allowed to move through distinct energy 
channels for a wide range of energy which is observed from the spectra
given in Fig.~\ref{updn}. The term $\vec{\mbox{\boldmath $h$}}_i.\vec{\sigma}$
in the TB Hamiltonian (Eq.~\ref{equ2}) is responsible for it and this 
channel separation suggests us to design the system as a spin filter which
we discuss in the forthcoming sub-section. Before that, here 
we explain the reason behind the channel separation and approximate
the magnitude of misalignment of two different energy bands for up and down
spin electrons. As already discussed, the transmission characteristic is 
the net effect of the combined system where MQW is coupled to the MQR. 
In absence of any external electric field both these two regions contribute 
to the transmission for their full allowed energy bands since under this 
condition all the energy levels are conducting in nature. But, as the 
electric field is switched on the energy eigenstates associated with the 
MQR start to localize and even for very weak electric field the 
contributions from these states almost cease to zero (which can be clearly 
visible from Fig.~\ref{upmodel2}. Then, the essential contribution comes 
only from the MQW. Thus, both the nature and width of the $T_{\sigma}$-$E$ 
spectrum are eventually be controlled by the electric 
field $\xi$. In order to understand precisely the role of $\xi$ in 
determining the widths of $T_{\sigma}$-$E$ spectrum we have to focus on
the nature of energy band widths of the individual systems i.e., MQR and
MQW, since depending on $\xi$ either one (for strong $\xi$) or both of them 
(for weak $\xi$) contribute to electronic transmission. It is well known 
that for an ordered one-dimensional non-magnetic tight-binding ring 
characterized by on-site potential $\epsilon$ (say) and nearest-neighbor 
hopping integral $t$ (say), the allowed energy band lies within the range 
$\epsilon-2t$ to $\epsilon+2t$. Similar energy band is also obtained for 
an infinite one-dimensional perfect chain characterized by these parameters. 
Using this analogy we can figure out the energy band widths and also the
widths of $T_{\sigma}$-$E$ spectra for the sub-systems MQR and MQW 
including the combined system within the electrodes. To do this we start 
with the term $\vec{\mbox{\boldmath $h$}}_i.\vec{\sigma}$ which becomes
$h_z.\sigma_z$, since in our formulation we assume that all the magnetic
moments are equal in magnitude ($h_i=h$ (say) for all $i$) and they are
aligned along the positive z-axis. Now, at $\xi=0$, $\epsilon_i^r=0$ for
all $i$ of the MQR which results a perfect ring, While, the other part 
i.e., MQW always behaves like a perfect wire irrespective of $\xi$. This
simplification helps us to predict the energy band widths of the sub-systems
as follows. For the MQR the range of up spin
band is: $-h-2t_r$ to $-h+2t_r$ and for down spin it is: $h-2t_r$ to 
$h+2t_r$. While for the MQW, these are approximately as: $-h-2t_w$ to 
$-h+2t_w$ and $h-2t_w$ to $h+2t_w$, respectively. Therefore, for the chosen
set of parameter values the up spin bands for the individual geometries
lie within the range $-3$ eV to $1$ eV, and the range becomes $-1$ eV to
$3$ eV for the down spin bands. When these two sub-systems i.e., MQR and 
MQW couple to each other (by the coupling parameter $\lambda$ which is 
fixed at $1$ eV) to form a combined system, the above energy bands shift 
a very little and it results a net energy shift $\sim 2h$ ($=2$ eV).
This is exactly reflected in the $T_{\sigma}$-$E$ spectra.
Certainly, for the non-zero electric field the allowed energy bands get 
shifted, but then the essential contribution to the electronic transmission
comes from the MQW only which results a separation of the order of $2h$
i.e., $2$ eV between the $T_{\uparrow}$-$E$ and $T_{\downarrow}$-$E$.

\subsection{Spin dependent currents and spin polarization coefficient
for different system sizes}

Now, we turn to analyze the variation of spin dependent currents together 
\begin{figure}[ht]
{\centering \resizebox*{8cm}{6.5cm}{\includegraphics{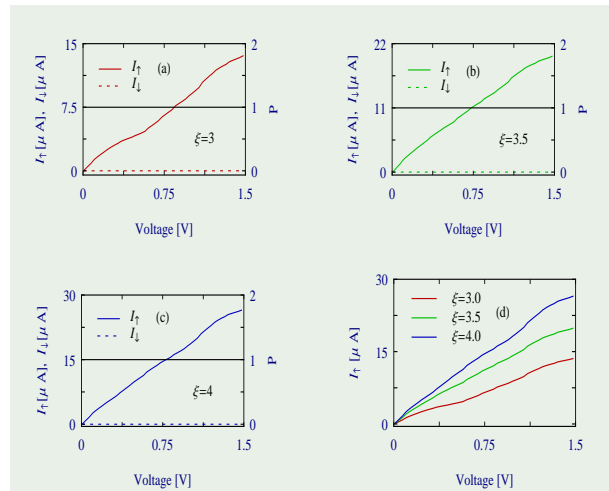}}\par}
\caption{(Color online). (a)-(c): Voltage dependence of up ($I_{\uparrow}$, 
left axis) and down ($I_{\downarrow}$, left axis) spin currents together 
with spin polarization coefficient ($P$, right axis, shown by the black curve) 
for different values of dimensionless electric field strength $\xi$ when the 
Fermi energy is fixed at $E_F=-1.75$ eV. The up spin currents shown 
in (a)-(c) are placed together in (d) for a better comparison of the 
amplitudes at different field strengths. The other physical parameters 
are: $N=42$, $M=20$, $m=32$, $n=33$, $p=10$, $q=11$ and $\lambda=1$ eV.}
\label{upcurrN42}
\end{figure}
with spin polarization coefficient and the role of external electric field
on them for different sizes of the MQR and MQW. With these characteristics 
the basic features of electron transmission can be understood in a much 
deeper way.

As illustrative example, in Fig.~\ref{upcurrN42} we plot the spin dependent 
\begin{figure}[ht]
{\centering \resizebox*{8cm}{6.5cm}{\includegraphics{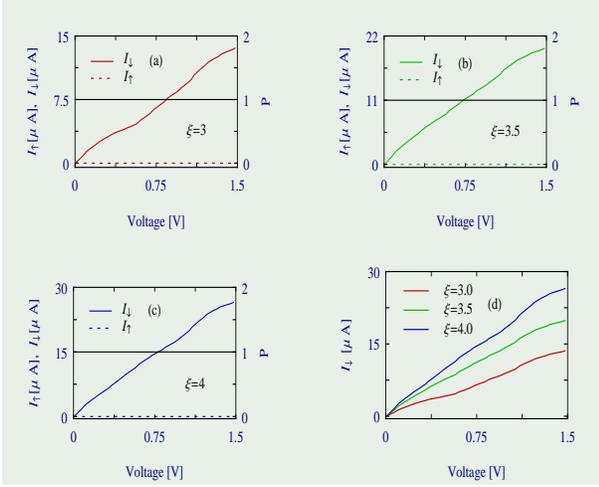}}\par}
\caption{(Color online). (a)-(c): Voltage dependence of up ($I_{\uparrow}$,
left axis) and down ($I_{\downarrow}$, left axis) spin currents together
with spin polarization coefficient ($P$, right axis, shown by the black curve) 
for different values of $\xi$ considering $E_F=1.75$ eV. The down spin 
currents shown in (a)-(c) are framed together in (d) for a better 
comparison of the amplitudes at different field strengths. All the other 
physical parameters remain exactly identical as taken in 
Fig.~\ref{upcurrN42}.}
\label{dncurrN42}
\end{figure}
\begin{figure}[ht]
{\centering \resizebox*{8cm}{6.5cm}{\includegraphics{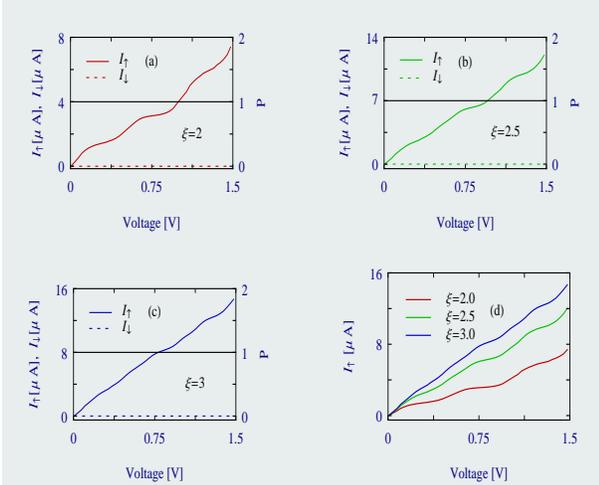}}\par}
\caption{(Color online). (a)-(c): $I_{\uparrow}$ (left axis), 
$I_{\downarrow}$ (left axis) and $P$ (right axis, shown by the black line) 
as a function of bias voltage $V$ for different values of $\xi$ taking 
$E_F=-1.75$ eV. The up spin currents shown in (a)-(c) are put together in 
(d) to compare the current amplitudes properly. The other parameters are: 
$N=100$, $M=40$, $m=75$, $n=76$, $p=20$, $q=21$ and $\lambda=1$ eV.}
\label{upcurrN100}
\end{figure}
currents $I_{\uparrow}$ and $I_{\downarrow}$, and spin polarization
coefficient $P$ as a function of applied bias voltage for different field
strengths when the Fermi energy is kept fixed at $E_F=-1.75$ eV. The results
computed for three distinct values of dimensionless electric field 
strength $\xi$ are shown in (a)-(c), and finally, the up spin currents 
presented in these three spectra are placed together in (d) to compare 
their amplitudes properly at different field strengths.
From the spectra it is observed that the current for down spin electrons
drops exactly to zero (dotted curve) for the entire voltage region, while
a finite current (solid curve) is obtained for the other spin electrons.
It reveals that electrons with only up spin are allowed to move from the
source to drain through the conductor, whereas down spin electrons are
totally blocked. The reason is that, setting the Fermi energy at 
$E_F=-1.75$ eV when we apply bias voltage only up spin channels appear 
within the voltage
window and they contribute to the current, but no conducting channel for
down spin electrons is available which yields a vanishing down spin current.
This phenomenon leads to the possibility of getting spin filtering action 
\begin{figure}[ht]
{\centering \resizebox*{8cm}{6.5cm}{\includegraphics{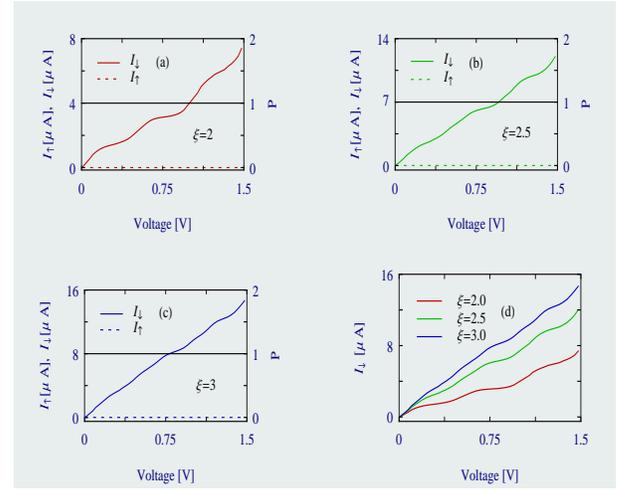}}\par}
\caption{(Color online). (a)-(c): $I_{\uparrow}$ (left axis),
$I_{\downarrow}$ (left axis) and $P$ (right axis, shown by the black line) 
for different values of $\xi$ setting $E_F=1.75$ eV. For a better comparison
of current amplitudes, the down spin currents shown in (a)-(c) are put 
together in (d). All the other physical parameters kept unchanged as taken 
in Fig.~\ref{upcurrN100}.}
\label{dncurrN100}
\end{figure}
using this bridge setup. The efficiency of spin filtration is depicted by
the polarization curve which shows $P=1$ throughout the bias window. 
This is expected since for the bias window down spin 
current ceases exactly to zero, while finite up spin current is obtained 
which yields perfect spin polarization (as clearly seen from 
Eq.~\ref{equ7}). Thus, our proposed quantum system can be utilized as 
a perfect spin filter for a wide voltage window.

The effect of in-plane electric field on spin current is quite interesting. 
For a fixed conductor-to-electrode coupling, described by the physical 
parameter $t_c$, the up spin current is enhanced significantly with 
increasing the dimensionless field strength $\xi$ (see 
Fig.~\ref{upcurrN42}(d)). This enhancement of current amplitude can be 
attributed following the transmittance-energy spectra (left column of 
Fig.~\ref{updn}) since current
is evaluated by integrating the transmission function (Eq.~\ref{equ6}).
The area under the transmission curve gets increased with the field
strength which results larger current across the bridge system. Usually, 
the current enhancement takes place by the coupling parameter $t_c$
in any bridge system~\cite{datta1,datta2,skm1,skm2}, but in our setup 
\begin{figure}[ht]
{\centering \resizebox*{8cm}{5cm}{\includegraphics{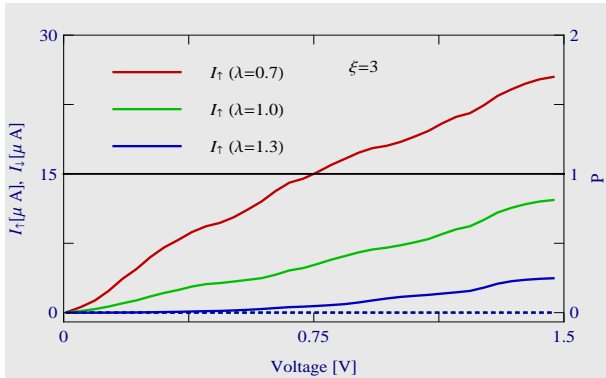}}\par}
\caption{(Color online). Voltage dependence of up ($I_{\uparrow}$, left axis) 
and down ($I_{\downarrow}$, left axis, dotted line) spin currents together 
with spin polarization coefficient ($P$, right axis) for different 
values of the ring to wire coupling strength $\lambda$ when $E_F=-1.85$ eV. 
The other parameters are: $N=150$, $M=50$, $m=113$, $n=114$, $p=25$ and 
$q=26$.}
\label{lambdaup}
\end{figure}
\begin{figure}[ht]
{\centering \resizebox*{8cm}{5cm}{\includegraphics{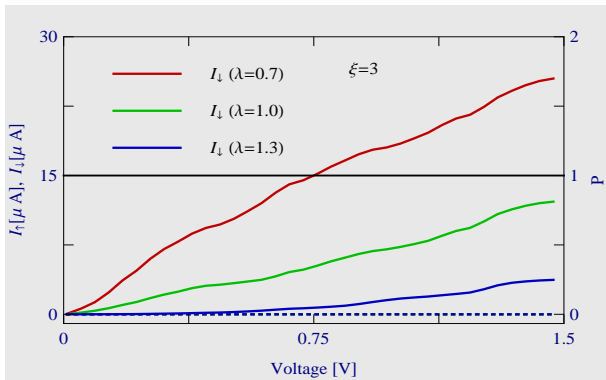}}\par}
\caption{(Color online). Voltage dependence of up ($I_{\uparrow}$, left axis) 
and down ($I_{\downarrow}$, left axis, dotted line) spin currents together 
with spin polarization coefficient ($P$, right axis) for different values 
of $\lambda$ setting $E_F=1.85$ eV. The other parameters are same as taken
in Fig.~\ref{lambdaup}.}
\label{lambdadn}
\end{figure}
we perform it {\em externally} with the help of in-plane electric field 
without directly changing other physical parameters of the system. It 
emphasizes that the presented system can be utilized as an {\em externally 
controlled spin based quantum device.}

An exactly similar behavior is also obtained for the down spin electrons 
when we set the Fermi energy $E_F=1.75$ eV. The variation of up and down 
spin currents along with the spin polarization coefficient are presented 
in Fig.~\ref{dncurrN42} considering the identical parameter values as taken 
in Fig.~\ref{upcurrN42}. From the spectra illustrated in 
Figs.~\ref{upcurrN42} and \ref{dncurrN42} we can predict that by tuning 
the Fermi energy to a suitable energy zone selective spin transfer can 
be achieved through our proposed two-terminal bridge setup.

The characteristic features of spin resolved currents ($I_{\uparrow}$ and
$I_{\downarrow}$) including the spin polarization coefficient ($P$) as 
a function of external bias for other system sizes of the MQR and MQW are
qualitatively similar to those with the bridge setup where $N$ and $M$ are
fixed at $42$ and $20$, respectively (Figs.~\ref{upcurrN42} and 
\ref{dncurrN42}). The results are presented in Figs.~\ref{upcurrN100}
and \ref{dncurrN100} for different strengths of the dimensionless electric
field $\xi$ and they are worked out for $N=100$ and $M=40$. Observations
of these spin dependent currents together with spin polarization for 
different system sizes (see Figs.~\ref{upcurrN42}-\ref{dncurrN100}) 
clearly suggest that the results are quite robust, and thus, can be 
utilized to achieve spin selective currents as well as high degree of
spin polarization in a two-terminal geometry.

\subsection{Effect of $\lambda$ on spin currents and spin polarization 
coefficient}

In order to elucidate the role played by the ring-to-wire coupling 
$\lambda$ on spin polarization and spin selective transmissions, in 
Figs.~\ref{lambdaup} and \ref{lambdadn} we present the results for a 
bridge setup with $N=150$ and $M=50$ considering different values of 
$\lambda$. In Fig.~\ref{lambdaup} the results are shown when the Fermi 
energy is fixed at $E_F=-1.85$ eV, while it is $1.85$ eV for the other 
figure (Fig.~\ref{lambdadn}). From the spectra it is observed that the 
selective spin current (up or down), associated with the choice of Fermi 
energy, gradually decreases with increasing the strength $\lambda$.
In presence of finite electric field, the coupling between the ordered 
(MQW) and disordered (MQR) regions gets enhanced with increasing the 
coupling parameter $\lambda$. Therefore, the ordered states generated
from the MQW are more affected by the disordered states appearing from the
MQR for higher $\lambda$ which results lower current. All the other 
characteristic features remain qualitatively similar to those as discussed
in the previous sub-section.

\subsection{Practicability consideration: Comparison of spin
current amplitudes for the zero and non-zero electric field cases}

\begin{figure}[ht]
{\centering \resizebox*{8cm}{5cm}{\includegraphics{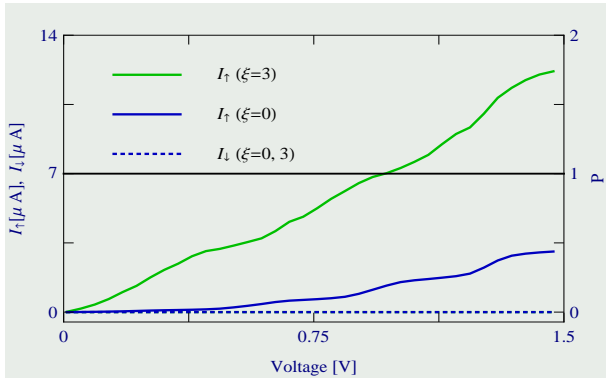}}\par}
\caption{(Color online). Comparison between up spin currents
(left axis) for zero and non-zero field cases is shown considering 
$E_F=-1.85$ eV. The variations of down spin currents (left axis) together
with spin polarization coefficient (right axis) are also presented. Here
we choose $N=150$, $M=50$, $m=113$, $n=114$, $p=25$, $q=26$ and 
$\lambda=1$ eV.}
\label{upzero}
\end{figure}

To demonstrate the crucial role of external electric field 
on regulation of spin current amplitude across the junction shown in
Fig.~\ref{model}, now it is interesting to compare spin dependent currents 
computed for zero and non-zero field cases.
First we focus on the results given in Fig.~\ref{upzero} where spin 
dependent currents are computed for two different field strengths, $\xi=0$
and $\xi=3$, setting the Fermi energy $E_F=-1.85$ eV. The results are
very significant. For $\xi=0$, the up spin current becomes two small
(solid blue line), while it rises to a large value for the non-zero 
field (green line). This enhancement of current amplitude can be justified
from our previous analysis. As noted in sub-section B, we see that in the
zero field limit both the MQR and MQW contribute to the current where
the transmission spectrum exhibits sharp resonant peaks which provide
a sufficiently small current upon integrating the transmission function.
On the other hand, for non-zero and moderate field strengths 
transmittance-energy spectrum looks like as obtained in a conventional
magnetic wire with broader resonant peaks which results a larger current
across the junction.
\begin{figure}[ht]
{\centering \resizebox*{8cm}{5cm}{\includegraphics{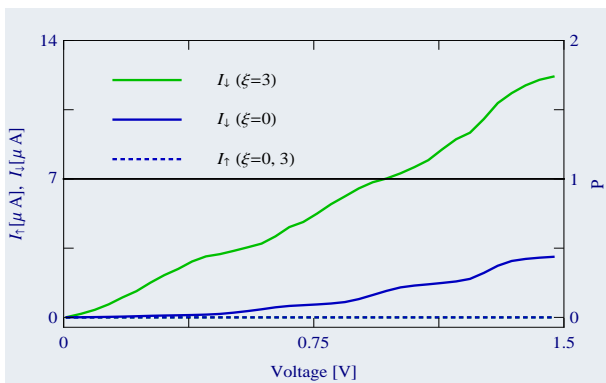}}\par}
\caption{(Color online). Comparison between down spin 
currents (left axis) for zero and non-zero field cases is shown considering
$E_F=1.85$ eV. The variations of up spin currents (left axis) together
with spin polarization coefficient (right axis) are also presented. All
the other parameters are same as in Fig.~\ref{upzero}.}
\label{dnzero}
\end{figure}
The nature of vanishing down spin currents and perfect 
spin polarization shown in this figure (Fig.~\ref{upzero}) can be easily 
understood from the earlier analysis.

Similar arguments are also given to explain the results plotted in 
Fig.~\ref{dnzero} where we set $E_F=1.85$ eV.

These observations (Figs.~\ref{upzero} and \ref{dnzero})
can be summarized by stating that the $I_{\sigma}$-$V$ behavior is highly
sensitive to the external electric field and can be utilized to design 
tailor made spintronic devices.

\subsection{Comparison of spin current magnitudes with other bridge setups}

\begin{figure}[ht]
{\centering \resizebox*{6.8cm}{3.5cm}{\includegraphics{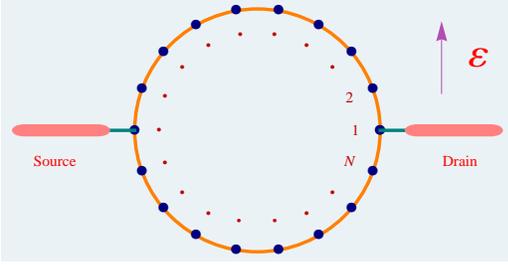}}\par}
\caption{(Color online). Another setup where MQR is no 
longer coupled to MQW like the previous two bridges. The ring, subjected 
to an electric field, is directly coupled to two NM electrodes.}
\label{model2}
\end{figure}

The results analyzed so far are worked out for the model
geometry shown in Fig.~\ref{model} where the electrodes are attached to
the MQW. Now, to inspect the pivotal role played by the MQW finally in 
this sub-section we present a comparative study of spin current magnitudes
\begin{figure}[ht]
{\centering \resizebox*{6.8cm}{4.5cm}{\includegraphics{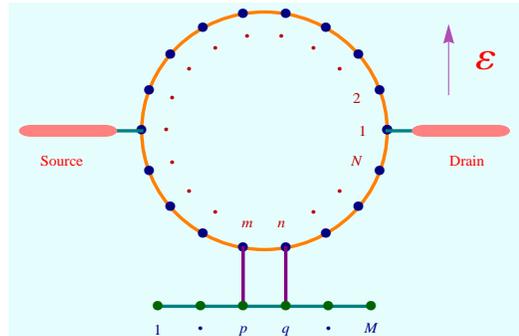}}\par}
\caption{(Color online). A different bridge configuration 
compared to Fig.~\ref{model}, where MQR is directly coupled to source and 
drain electrodes. The setup within the electrodes remains unaltered as 
considered in Fig.~\ref{model}.}
\label{model1}
\end{figure}
\begin{figure}[ht]
{\centering \resizebox*{8cm}{5cm}{\includegraphics{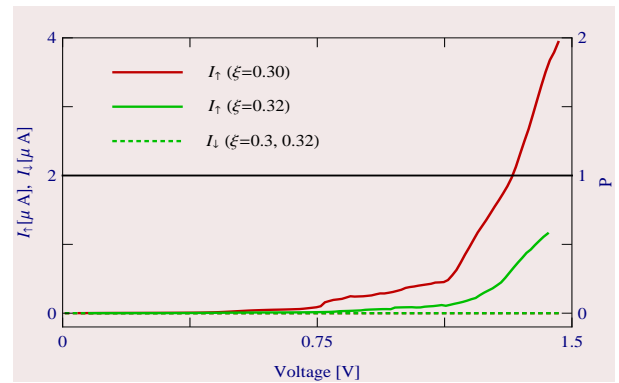}}\par}
\caption{(Color online). Voltage dependence of up 
($I_{\uparrow}$, left axis) and down ($I_{\downarrow}$, left axis) spin 
currents together with spin polarization coefficient ($P$, right axis) for 
different values of $\xi$ considering $E_F=-1.75$ eV for the setup given 
in Fig.~\ref{model1}. Here we choose $N=42$, $M=20$, $m=32$, $n=33$, $p=10$,
$q=11$ and $\lambda=1$ eV.}
\label{upmodel1}
\end{figure}
considering two other different bridge setups with respect to 
Fig.~\ref{model}. They are schematically shown in Figs.~\ref{model1} and 
\ref{model2}, respectively. In one case, the source and drain are attached 
to the MQR (see Fig.~\ref{model1}), instead of the MQW, and within these 
electrodes the setup remains unchanged as taken in Fig.~\ref{model}. While, 
in the other case only the MQR is taken into account 
within the electrodes (see Fig.~\ref{model2}) to form a simple two-terminal 
bridge setup. Now we describe the results for these setups one by one.

\begin{figure}[ht]
{\centering \resizebox*{8cm}{5cm}{\includegraphics{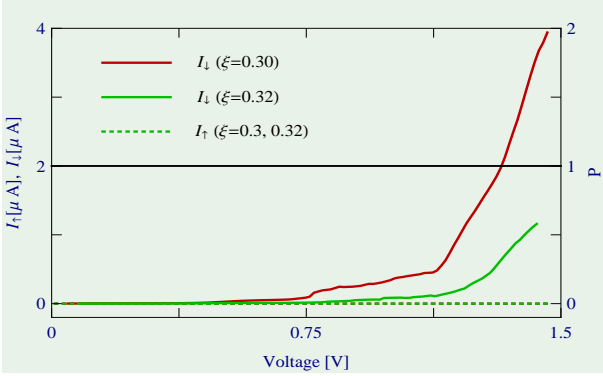}}\par}
\caption{(Color online). Voltage dependence of up 
($I_{\uparrow}$, left axis) and down ($I_{\downarrow}$, left axis) spin 
currents along with spin polarization coefficient ($P$, right axis) for 
different values of $\xi$ taking $E_F=1.75$ eV for the identical setup 
considered in Fig.~\ref{upmodel1}. The parameters are same as taken 
in Fig.~\ref{upmodel1}.}
\label{dnmodel1}
\end{figure}

In Figs.~\ref{upmodel1} and \ref{dnmodel1} the variation 
of spin dependent currents ($I_{\uparrow}$ and $I_{\downarrow}$) along with 
spin polarization coefficient $P$ as a function of applied bias voltage $V$
are shown for two distinct values of the field strength 
$\xi$. Focusing on the characteristics presented in the spectra 
(Figs.~\ref{upmodel1} and \ref{dnmodel1}), two observations are noteworthy. 
First, the current amplitudes for non-zero fields are too small compared 
to the current obtained in the model Fig.~\ref{model}. Second, even for 
a slight increment of field strength $\xi$ current amplitude reduces very 
sharply, unlike the initial configuration i.e., Fig.~\ref{model} where 
current amplitude
gets increased with increasing $\xi$. These features can 
be explained as follows. As stated, the MQR behaves like a correlated 
disordered system in presence of non-zero field $\xi$ since 
$\epsilon_i^r$'s are now field dependent, and therefore, the combined system
can be regarded as an ordered-disordered coupled system. Thus, in the bridge
given in Fig.~\ref{model1}, an electron which is coming from the source gets 
injected into the disordered region (MQR) and after traversing throughout 
the material (MQR and MQW) it eventually leaves from the disordered part
(MQR) to enter into the drain. The width of disorderness becomes wider
with the field strength $\xi$ and hence the energy levels associated with 
the MQR become more localized which lead to the reduced current across
the junction. Comparing the results shown in Figs.~\ref{upmodel1} and 
\ref{dnmodel1}, it is observed that the current amplitude decreases
significantly even for a small increment of $\xi$ (from $0.3$ to $0.32$),
and if we increase $\xi$ further current practically disappears. This 
\begin{figure}[ht]
{\centering \resizebox*{8cm}{5cm}{\includegraphics{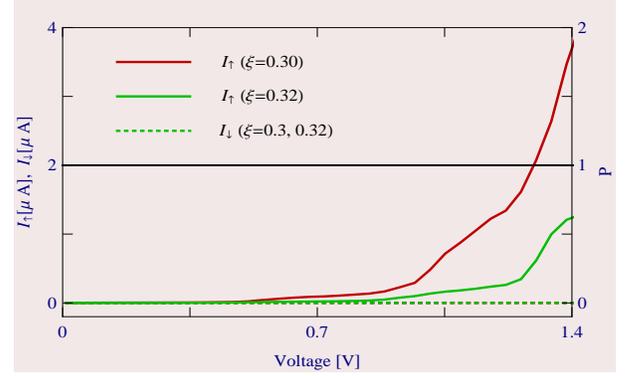}}\par}
\caption{(Color online). $I_{\uparrow}$ (left axis), 
$I_{\downarrow}$ (left axis) and $P$ (right axis) as a function of voltage
$V$ for different values of $\xi$ considering $N=42$, $\lambda=1$ eV and 
$E_F=-1.75$ eV for the setup given in Fig.~\ref{model2}.}
\label{upmodel2}
\end{figure}
\begin{figure}[ht]
{\centering \resizebox*{8cm}{5cm}{\includegraphics{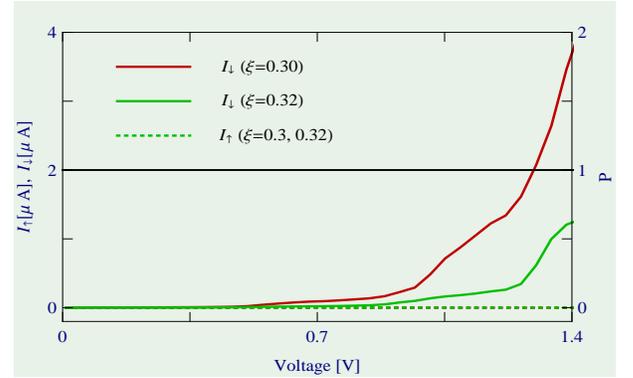}}\par}
\caption{(Color online). $I_{\uparrow}$ (left axis),
$I_{\downarrow}$ (left axis) and $P$ (right axis) as a function of voltage
$V$ for different values of $\xi$ considering $N=42$, $\lambda=1$ eV and
$E_F=1.75$ eV for the identical setup as taken in Fig.~\ref{upmodel2}.}
\label{dnmodel2}
\end{figure}
scenario is exactly opposite what we get in our previous geometry. In that
model (Fig.~\ref{model}) the ordered region (MQW) gradually decouples from
the localized region (MQR) with increasing $\xi$ and the probability of
traversing electrons through the ring also decreases which leads to enhanced
spin current. Thus, by tuning $\xi$ we eventually enhance the probability
of traversing electron through the wire which results larger current in
the junction (Fig.~\ref{model}), which is no longer possible if the 
electrodes are coupled directly to the ring (Fig.~\ref{model1}) instead 
of the wire as clearly seen from our results given in Figs.~\ref{upmodel1} 
and \ref{dnmodel1}.

The existence of MQW does not provide any new significant 
behavior on spin dependent currents when the electrodes are coupled to the 
MQR, since for such a configuration electrons are eventually entering into 
the drain from a correlated disordered region for non-zero $\xi$. To 
corroborate this fact, in Figs.~\ref{upmodel2} and \ref{dnmodel2} we 
present the behavior of spin dependent currents including spin polarization 
coefficient for the junction configuration given in Fig.~\ref{model2}, where 
the ring is not attached to any MQW. Comparing the results of 
Figs.~\ref{upmodel1}, \ref{dnmodel1}, \ref{upmodel2} and \ref{dnmodel2}, 
we predict that the current-voltage characteristics show very less 
sensitivity on the MQW when the electrodes are coupled to the MQR. Thus, 
in short, we can emphasize that to design a tailor made spin based quantum 
device the proposed quantum system given in Fig.~\ref{model} is the most 
suitable one.

\section{Concluding remarks}

To conclude, in the present work we address a new approach of getting
spin selective transmission through a non-magnetic -- magnetic -- non-magnetic 
bridge system based on Green's function formalism. The magnetic system 
consists of a quantum ring which is directly coupled to a quantum wire 
and subjected to an in-plane electric field. From our results we find that 
the transmission spectrum gets significantly influenced by the electric field
which directly reflects the current-voltage characteristics. Tuning the
Fermi energy to a suitable energy zone a high degree of spin polarization
($\sim 100 \%$) can also be achieved for a wide range of bias voltage for 
this setup. Our theoretical analysis promotes practical applications of
{\em externally controlled} spin polarized quantum devices.

All the results presented in this communication are worked out at 
absolute zero temperature though its finite temperature extension is quite
trivial. But, the thing is that at finite (low) temperatures no new 
phenomenon will appear since the thermal broadening of energy levels is
too weak compared to the energy level broadening caused by the coupling
of the bridging conductor to the side attached 
electrodes~\cite{datta1,datta2}.

Before we end, it should be noted that to investigate spin selective transfer 
through this two-terminal geometry we compute all the numerical results 
considering some typical values of the physical parameters. But, all the
physical phenomena studied here remain absolutely invariant for any other 
choices of the physical parameters describing the system. These features 
certainly demand the robustness of our analysis and give us confidence 
to propose an experiment in this line.


\begin{thebibliography}{99}

\bibitem{sp1} S. A. Wolf, D. D. Awschalom, R. A. Buhrman, J. M. Daughton,
S. von Moln\'{a}r, M. L. Roukes, A. Y. Chtchelkanova, and D. M. Treger,
Science \textbf{294}, 1488 (2001).

\bibitem{sp2} G. Prinz, Science \textbf{282}, 1660 (1998).

\bibitem{sp3} G. Prinz, Phys. Today \textbf{48}, 58 (1995).

\bibitem{nano1} J. Chen, M. A. Reed, A. M. Rawlett, and J. M. Tour,
Science \textbf{286}, 1550 (1999).

\bibitem{nano2} P. Ball, Nature (London) \textbf{404}, 918 (2000).

\bibitem{ex1} L. P. Rokhinson, V. Larkina, Y. B. Lyanda-Geller, L. N.
Pfeiffer, and K. W. West, Phys. Rev. Lett. \textbf{93}, 146601 (2004).

\bibitem{ex2} S. Sahoo, T. Kontos, J. Furer, C. Hoffmann, M. Gr\"{a}ber,
A. Cottet, and C. Sch\"{o}nenberger, Nature Phys. \textbf{1}, 99 (2005).

\bibitem{ex3} N. Tombros, C. Jozsa, M. Popinciuc, H. T. Jonkman,
and B. J. van Wees, Nature \textbf{448}, 571 (2007).

\bibitem{gmr} M. N. Baibich, J. M. Broto, A. Fert, F. N. Van Dau,
F. Petroff, P. Etienne, G. Creuzet, A. Friederich, and J. Chazelas,
Phys. Rev. Lett. \textbf{61}, 2472 (1998).

\bibitem{dev1} B. E. Kane, Nature (London) \textbf{393}, 133 (1998).

\bibitem{dev2} V. Privman, I. D. Vagner, and G. Kventsel, Phys. Lett. A
\textbf{239}, 141 (1998).

\bibitem{dev3} G. Burkard, D. Loss, and D. P. DiVincenzo, Phys. Rev. B
\textbf{59}, 2070 (1999).

\bibitem{dev4} Yu. V. Pershin, I. D. Vagner, and P. Wyder, J. Phys.:
Condens. Matter \textbf{15}, 997 (2003).

\bibitem{spinsteps1} J. St\"{o}hr and H. C. Siegmann, {\em Magnetism -- 
From Fundamental to Nanoscale Dynamics} (Springer, 2006).

\bibitem{spinsteps2} S. Maekawa and T. Shinjo, {\em Spin Dependent
Transport in Magnetic Nanostructures}, (CRC Press, 2002).

\bibitem{th1} H. Yin, T. L\"{u}, X. Liu, and H. Xue, Phys. Lett. A 
\textbf{285}, 373 (2009).

\bibitem{th2} F. Chi and S. Li, J. Appl. Phys. \textbf{100}, 113703 (2006).

\bibitem{sm1} M. Dey, S. K. Maiti, and S. N. Karmakar, Phys. Lett. A
\textbf{374}, 1522 (2010).

\bibitem{th3} M. W. Wu, J. Zhou, and Q. W. Shi, Appl. Phys. Lett. 
\textbf{6}, 85 (2004).

\bibitem{sm2} M. Dey, S. K. Maiti, and S. N. Karmakar, J. Comput.
Theor. Nanosci. \textbf{8}, 253 (2011).

\bibitem{th4} A. A. Shokri, M. Mardaani, and K. Esfarjani, Physica E 
\textbf{27}, 325 (2005).

\bibitem{thnew1} A. D. G\"{u}cl\"{u}, P. Potasz, and 
P. Hawrylak, Phys. Rev. B \textbf{84}, 035425 (2011).

\bibitem{thnew2} O. Voznyy, A. D. G\"{u}cl\"{u}, 
P. Potasz, and P. Hawrylak, Phys. Rev. B \textbf{83}, 165417 (2011).

\bibitem{thnew3} M. Modarresi, M. R. Roknabadi, and 
N. Shahtahmasebi, J. Magn. Magn. Mater. \textbf{350}, 6 (2014).

\bibitem{thnew4} K. Szalowski, J. Magn. Magn. Mater. 
\textbf{382}, 318 (2015). 

\bibitem{th5} M. Lee and C. Bruder, Phys. Rev. B \textbf{73}, 085315 (2006).

\bibitem{th6} J. H. Ojeda, M. Pacheco, and P. A. Orellana, Nanotechnology 
\textbf{20}, 434013 (2009).

\bibitem{sm3} M. Dey, S. K. Maiti, and S. N. Karmakar, Eur. Phys.
J. B \textbf{80}, 105 (2011).

\bibitem{th7} K. Chang and F. M. Peeters, Solid State Commun. \textbf{120}, 
181 (2001).

\bibitem{th8} A. A. Shokri and M. Mardaani, Solid State Commun. 
\textbf{137}, 53 (2006).

\bibitem{expt} D. Jin, Z. Li, M. Xiao, G. Jin, and A. Hu, J. Appl. 
Phys. \textbf{99}, 08T304 (2004).

\bibitem{trend1} W. Long, Q.-F. Sun, H. Guo, and J. Wang, Appl. Phys. Lett. 
\textbf{83}, 1397 (2003).

\bibitem{trend2} P. Zhang, Q. K. Xue, and X. C. Xie, Phys. Rev. Lett.
\textbf{91}, 196602 (2003).

\bibitem{intrinsic1} Q.-F. Sun and X. C. Xie, Phys. Rev. B \textbf{91},
235301 (2006).

\bibitem{intrinsic2} Q.-F. Sun and X. C. Xie, Phys. Rev. B \textbf{71},
155321 (2005).

\bibitem{intrinsic3} F. Chi, J. Zheng, and L. L. Sun, Appl. Phys. Lett.
\textbf{92}, 172104 (2008).

\bibitem{intrinsic4} T. P. Pareek, Phys. Rev. Lett. \textbf{92}, 076601 
(2004).

\bibitem{intrinsic5} W. J. Gong, Y. S. Zheng, and T. Q. L\"{u}, Appl. Phys. 
Lett. \textbf{92}, 042104 (2008).

\bibitem{intrinsic6} H. F. L\"{u} and Y. Guo, Appl. Phys. Lett. \textbf{91}, 
092128 (2007).

\bibitem{rashba} Y. A. Bychkov and E. I. Rashba, JETP Lett. \textbf{39},
78 (1984).

\bibitem{dressel} G. Dresselhaus, Phys. Rev. \textbf{100}, 580 (1955).

\bibitem{winkler} R. Winkler, {\em Spin-orbit coupling effects in
two-dimensional electron and hole Systems} (Springer, 2003).

\bibitem{skm} S. K. Maiti, S. Sil, and A. Chakrabarti, Phys. Lett. A
\textbf{376}, 2147 (2012).

\bibitem{gt1} G. Engels, J. Lange, Th. Sch\"{a}pers, and H. L\"{u}th,
Phys. Rev. B \textbf{55}, R1958 (1997).

\bibitem{gt2} L. Meier, G. Salis, I. Shorubalko, E. Gini, S. Sch\"{o}n,
and K. Ensslin, Nature Physics \textbf {3}, 650 (2007).

\bibitem{gt3} J. Premper, M. Trautmann, J. Henk, and P. Bruno, Phys.
Rev. B \textbf{76}, 073310 (2007).

\bibitem{gt4} C.-M. Hu, J. Nitta, T. Akazaki, H. Takayanagi, J. Osaka,
P. Pfeffer, and W. Zawadzki, Phys. Rev. B \textbf{60}, 7736 (1999).

\bibitem{gt5} D. Grundler, Phys. Rev. Lett. \textbf{84}, 6074 (2000).

\bibitem{mlt1} P. F\"{o}ldi, O. K\'{a}lm\'{a}n, M. G. Benedict, and
F. M. Peeters, Phys. Rev. B \textbf{73}, 155325 (2006).

\bibitem{mlt2} A. A. Kislev and K. W. Kim, J. App. Phys. \textbf{94}, 4001
(2003).

\bibitem{mlt3} S. Souma and B. K. Nikoli\'{c}, Phys. Rev. B \textbf{70},
195346 (2004).

\bibitem{mlt4} B. K. Nikoli\'{c} and S. Souma, Phys. Rev. B \textbf{71},
195328 (2005).

\bibitem{mlt5} M. Dey, S. K. Maiti, S. Sil, and S. N. Karmakar, J. Appl.
Phys. \textbf{114}, 164318 (2013).

\bibitem{skm3} S. K. Maiti, Phys. Lett. A \textbf{379}, 361 (2015).

\bibitem{raba} G. Cohen, O. Hod, and E. Rabani, Phys. Rev. B \textbf{76},
235120 (2007).

\bibitem{smpola} M. Dey, S. K. Maiti, and S. N. Karmakar, J. Appl. Phys.
\textbf{109}, 024304 (2011).

\bibitem{tb1} P. Orellana and F. Claro, Phys. Rev. Lett. \textbf{90},
178302 (2003).

\bibitem{tb2} G. Stefanucci, E. Perfetto, S. Bellucci, and M. Cini,
Phys. Rev. B \textbf{79}, 073406 (2009).

\bibitem{tb3} A.-M. Guo and Q.-F. Sun, Phys. Rev. Lett. \textbf{108},
218102 (2012).

\bibitem{tb4} P. A. Orellana, M. L. Ladr\'{o}n de Guevara, M. Pacheco,
and A. Latg\'{e}, Phys. Rev. B \textbf{68}, 195321 (2003).

\bibitem{tb5} M. Modarresi, M. R. Roknabadi, and N. Shahtahmassebi,
Physica B \textbf{415}, 62 (2013).

\bibitem{tb6} M. Modarresi, M. R. Roknabadi, N. Shahtahmassebi, D. Vahedi,
and H. Arabshahi, Physica E \textbf{43}, 402 (2010).

\bibitem{tb7} S. K. Maiti, Phys. Lett. A \textbf{366}, 114 (2007).

\bibitem{tb8} S. Sil, S. K. Maiti, and A. Chakrabarti, Phys. Rev. Lett.
\textbf{101}, 076803 (2008).

\bibitem{datta1} S. Datta, {\em Electronic transport in mesoscopic systems}
(Cambridge University Press, Cambridge, 1995).

\bibitem{datta2} S. Datta, {\em Quantum transport: Atom to transistor}
(Cambridge University Press, Cambridge, 2005).

\bibitem{rai} D. Rai and M. Galperin, Phys. Rev. B \textbf{86}, 045420 
(2012).

\bibitem{nmn} R. Naaman and D. H. Waldeck, J. Phys. Chem. Lett. \textbf{3},
2178 (2012).

\bibitem{skm1} M. Dey, S. K. Maiti, and S. N. Karmakar, Org. Electron.
\textbf{12}, 1017 (2011).

\bibitem{skm2} P. Dutta, S. K. Maiti, and S. N. Karmakar, Org. Electron.
\textbf{11}, 1120 (2010).

\end{thebibliography}
\end{document}